\documentclass[prb,twocolumn,showpacs,floatfix]{revtex4}
\usepackage{graphicx}
\usepackage{dcolumn}
\usepackage{bm}
\begin{document}
\title{Phonon-Magnon coupling in CoF$_2$ investigated by time-of-flight neutron spectroscopy}
\author{Tapan Chatterji$^1$, Mohamed Zbiri$^1$, Stephane Rols$^1$}
\address{$^1$Institut Laue-Langevin, B.P. 156, 38042 Grenoble Cedex 9, France\\
}
\date{\today}

\begin{abstract}
We report the results of inelastic neutron scattering investigation on the model antiferromagnet CoF$_2$ by time-of-flight neutron spectroscopy. We measured the details of the scattering function $S({\bf Q},\omega)$ as a function of temperature with two different incident neutron wavelengths. The temperature and Q dependence of the measured scattering function suggests the presence of magnon-phonon coupling in almost all branches. The present results are in agreement with the strong magnetoelastic effects observed previously.

\end{abstract}
\pacs{75.25.+z}
\maketitle
The lattice dynamics of a crystal is normally assumed to be independent of the spin dynamics of the magnetic moments associated with the atoms and ions of the crystal. Similarly while describing the motions of the spins in a magnetically ordered systems it is usual to assume that the equilibrium positions of the magnetic ions are frozen. The approximation gives a fairly well description of the magnetic system and is in the same spirit as the adiabatic or Born-Oppenheimer approximation which is normally used for band structure calculations. In this approximation the  phonon and magnons are separate entities with their own dispersion relations and the Hamiltonian of the system separates out into a lattice and a magnetic part. In reality however there always exists a coupling term that involves displacements both in the positions of the atoms in the crystal and in the orientations of the spin vectors associated with these atoms. Thus the collective excitations are neither pure lattice waves nor spin waves but rather magnetoelastic waves. Such magnetoelastic waves are predicted to exist theoretically long time ago \cite{akhiezer68}. With the advent of enormous progress in first-principle calculations it is now possible to do phonon calculations of magnetic solids in the magnetically ordered phase by taking spin degrees of freedom into account. The difference between the results of the calculations done by taking and not taking magnetic degrees of freedom into account can be often very significant as has been amply demonstrated by such recent calculations on Fe-arsenide compounds \cite{zbiri10}. It is possible to study magnetoelastic effects experimentally by using neutron scattering techniques. Our recent neutron diffraction investigations \cite{chatterji10a,chatterji10b} have shown strong static magnetoelastic coupling in transition metal difluorides MnF$_2$, FeF$_2$, CoF$_2$ and NiF$_2$. There exist indirect evidences for dynamic magnetoelastic or phonon-magnon coupling in FeF$_2$, CoF$_2$ and NiF$_2$ \cite{macfarlane71,hutchings70}. However, a systematic study of the phonon-magnon coupling in this relatively simple system is still missing.

\begin{figure}
\resizebox{0.5\textwidth}{!}{\includegraphics{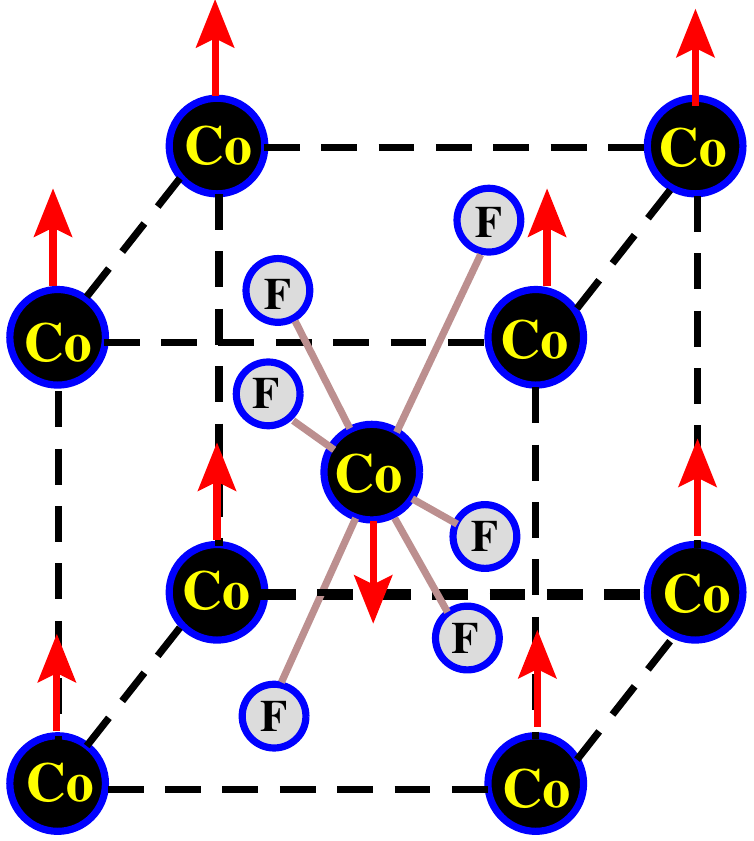}}
\caption {The antiferromagnetic structure adopted by $3d$ 
transition metal diflurides viz. CoF$_2$ with the rutile type crystal structure. 
The Co and F ions are labeled. The arrows on the Co represent
moment direction of the transition metal ions below the N\'eel temperature \cite{chatterji06}.
             } 
\label{structure}
\end{figure}

\begin{figure}
\resizebox{0.5\textwidth}{!}{\includegraphics{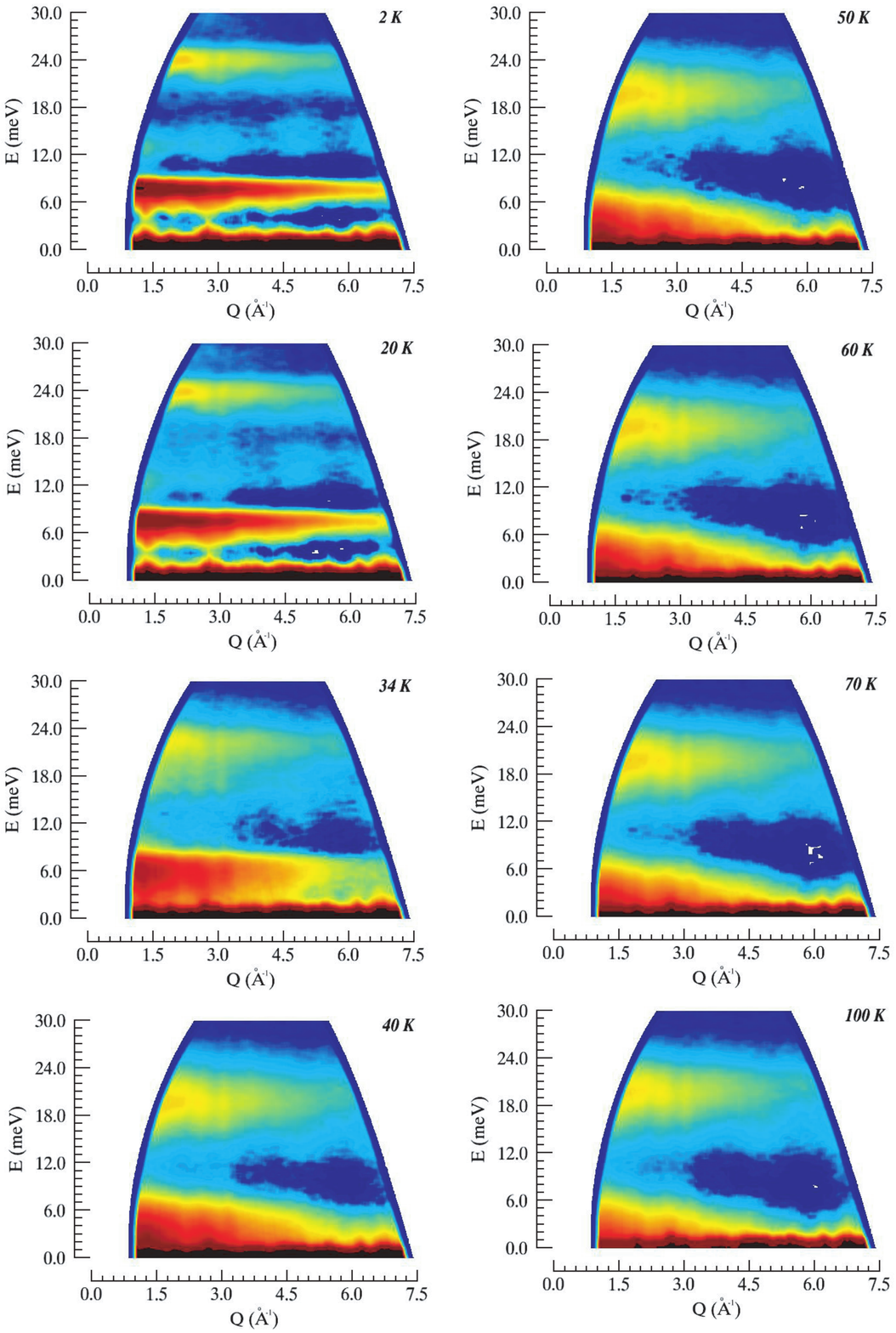}}
\caption {(Color online) Temperature evolution of the scattering function $S(Q,\omega)$ measured from CoF2 using incident neutron wavelength of $\lambda = 1.5 {\AA}$.} 
\label{spectra1}
\end{figure}

\begin{figure}
\resizebox{0.5\textwidth}{!}{\includegraphics{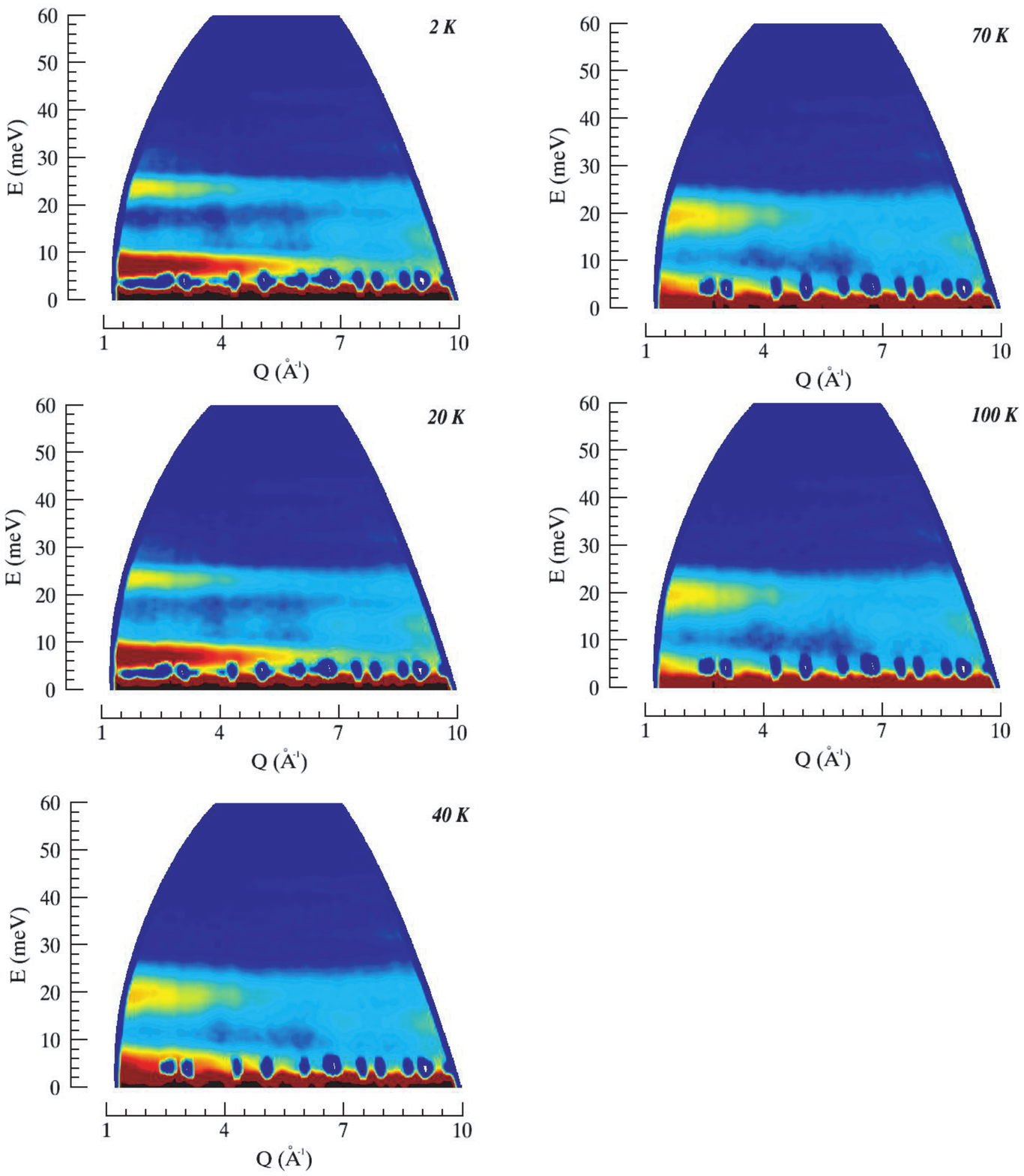}}
\caption {(Color online) Temperature evolution of the scattering function $S(Q,\omega)$ from CoF2 measured with the incident neutron wavelength of $\lambda = 1.1 {\AA}$.}
\label{spectra2}
\end{figure}

\begin{figure}
\resizebox{0.5\textwidth}{!}{\includegraphics{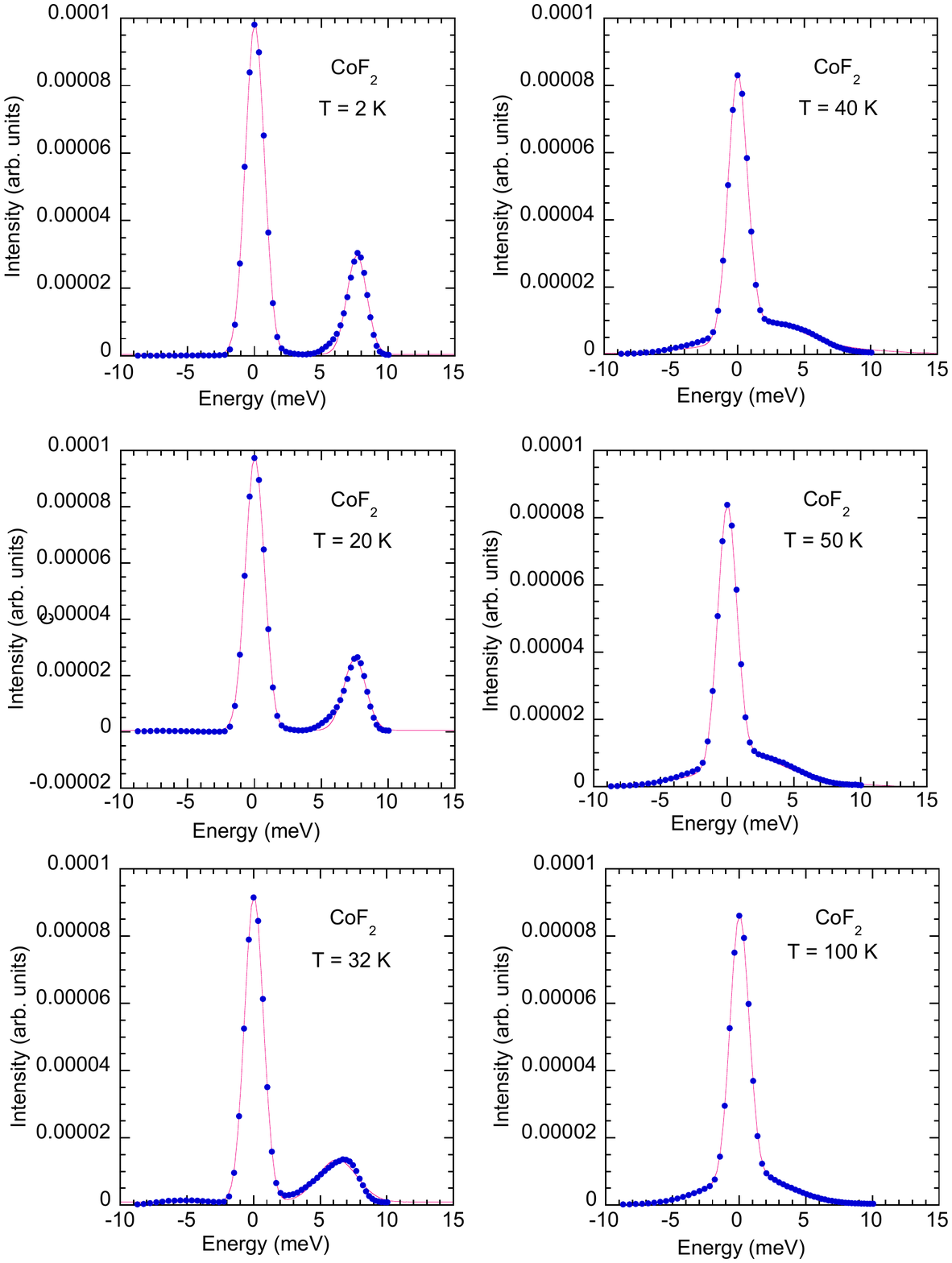}}

 \caption {Results of the fit of the low-Q data (integrated over Q in the range from 0.9 to 3.56 {\AA}$^{-1}$) from CoF$_2$ in the low-energy
   range ($\lambda_{i}$~= 1.5 {\AA}). Two Gaussian peaks have been fitted to the elastic peak at zero energy and the inelastic peak which is at about $E = 8$ meV at $T = 2$ K. The energy of inelastic peak decreases with increasing temperature and becomes zero at $T_N = 38$ K giving rise to quasielastic scattering which has been fitted with a Lorentzian function.}
\label{coffit1}
\end{figure}

 \begin{figure}
\resizebox{0.5\textwidth}{!}{\includegraphics{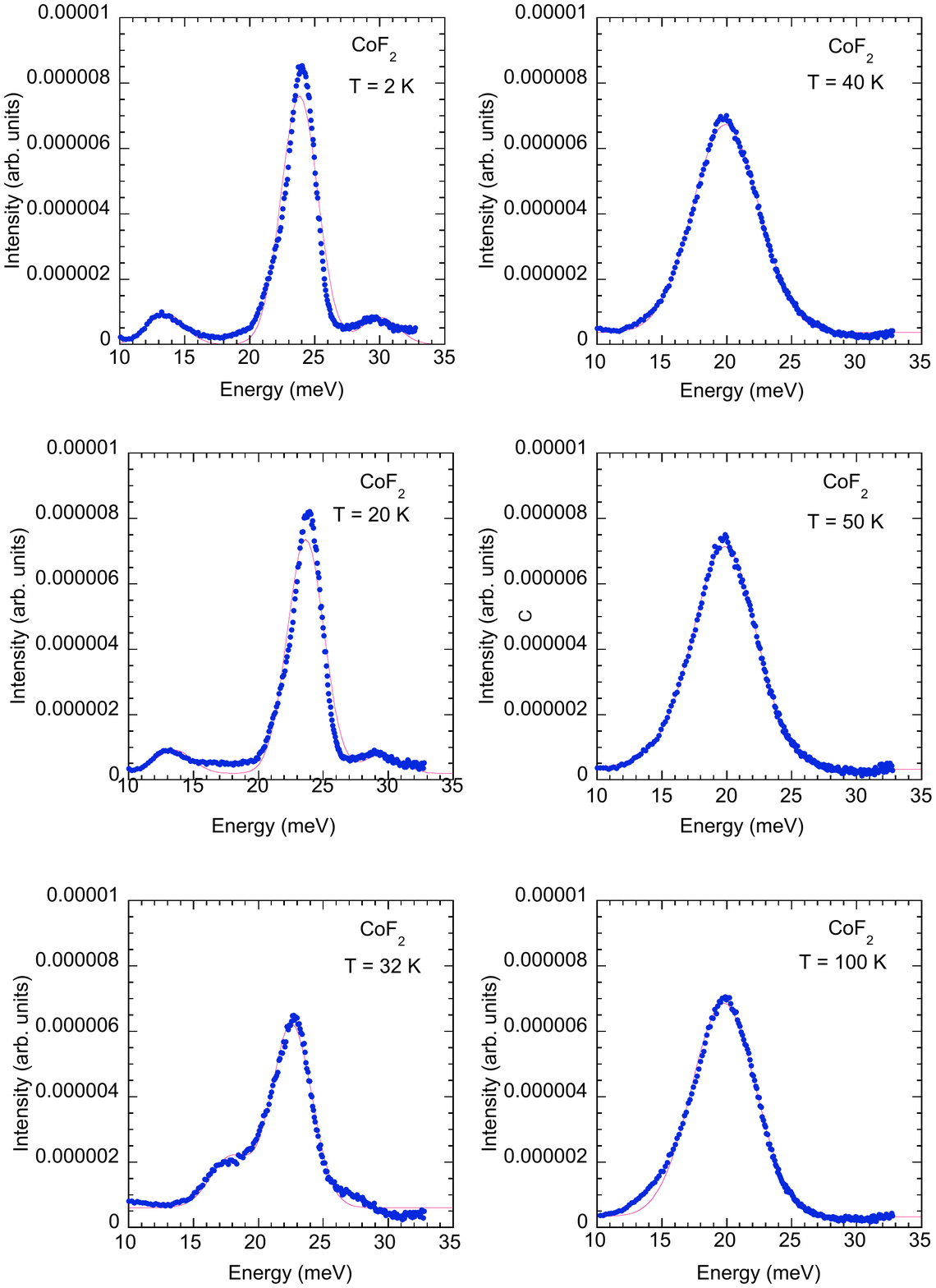}}
\caption {Results of the fit of the low-Q data (integrated over Q in the range from 0.9 to 3.56 {\AA}$^{-1}$) from CoF$_2$ in the
  higher-energy range ($\lambda_{i}$~= 1.5 {\AA}). At low temperatures below $T_N = 38$ K,  three peaks could be identified and they were fitted with three Gaussian functions. At higher temperature only a broad peak could be identified and has feen fitted with a single Gaussian function.  }
\label{coffit2}
\end{figure}

\begin{figure}
\resizebox{0.5\textwidth}{!}{\includegraphics{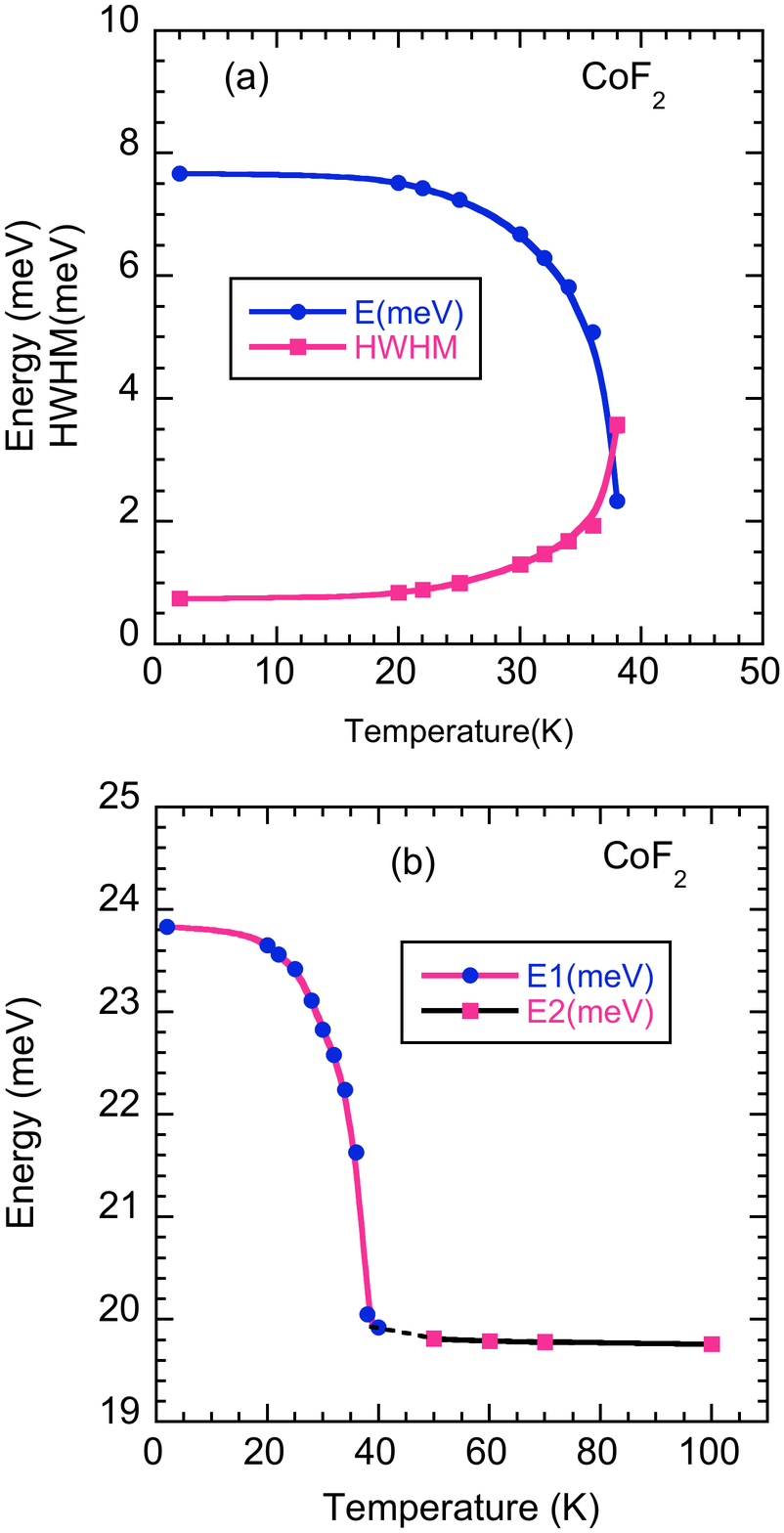}}
\caption {(Color online) (a)Temperature variation of the energy and the
  half-width at half-maximum (HWHM) of the low-energy inelastic magnetic peak
  of CoF$_2$. (b)Temperature variation of the energy of the high-energy
  inelastic magnetic peak of CoF$_2$ ($\lambda_{i}$~= 1.5 {\AA}).  }
\label{Tdep}
\end{figure}

\begin{figure}
\resizebox{0.5\textwidth}{!}{\includegraphics{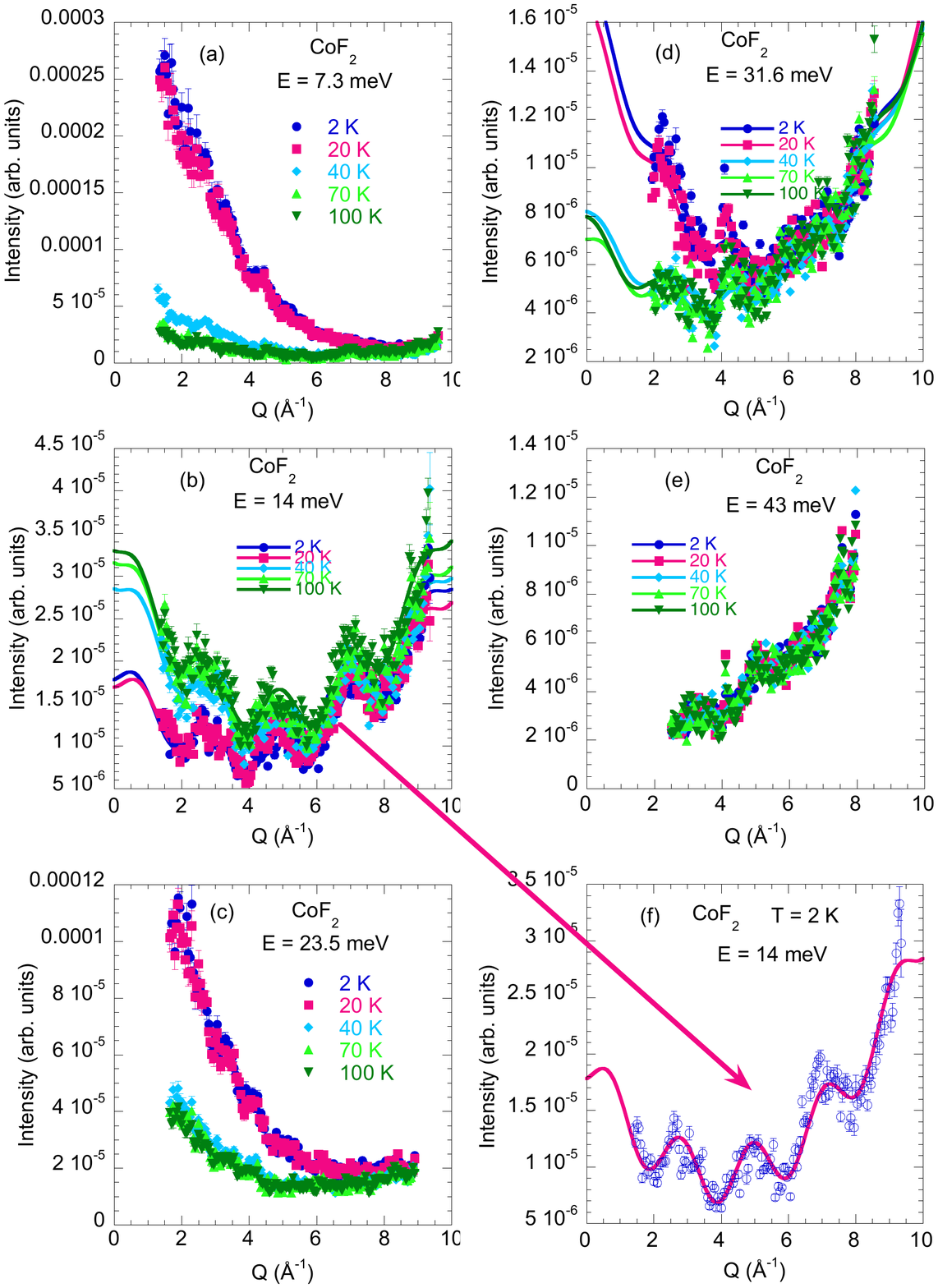}}
\caption {Q dependence of the intensity for the  inelastic peaks at $E = 7.3,
  14.0$, $23.5$, $31.6$ and $43.0$ meV at T = 2, 20, 40, 70 and  100 K ($\lambda_{i}$~= 1.1 {\AA}).}
\label{Qdep}
\end{figure}

CoF$_2$ belongs to the family of transition-metal difluoride, which has been the subject of intensive investigations \cite{staut49,ohlmann61,keffer52,nakamura55,stout53,moriya59,jaccarino59,jaccarino59a}.  
CoF$_2$, along with other transition-metal oxides MnF$_2$, FeF$_2$ and NiF$_2$, crystallize with the tetragonal  rutile-type structure in the $P4_2/mnm$
space group. However the magnetic properties of CoF$_2$ are more complex than those of isomorphous MnF$_2$, because the Co ion has unpaired angular momentum that plays an important role in determining its magnetic properties. CoF$_2$ orders \cite{erickson53,martel68,cowley73,strempfer04,jauch04} 
below $T_N \approx 39$ K with an antiferromagnetic structure \cite{chatterji06}, shown in Fig. \ref{structure}, with the propagation vector ${\bf k} = 0$.
The magnetic moments of Co ions at the corner  $(000)$ positions of the tetragonal unit cell are all parallel to the c-axis whereas those of the Co ions  at the $(\frac{1}{2}\frac{1}{2}\frac{1}{2})$ positions are oppositely oriented. 
 The presence of an orbital moment in CoF$_2$ makes it particularly interesting for the study of magnetoelastic coupling in this compound. We recently investigated \cite{chatterji09b} the hyperfine interaction in CoF$_2$ by high resolution neutron spectroscopy and concluded that the presence of the unquenched orbital moment of the Co ion in CoF$_2$ leads to its anomalous behavior compared to that of other Co-based compounds.
    
We have performed inelastic neutron scattering investigations on CoF$_2$ using the thermal time-of-flight neutron spectrometer IN4C at the Institute Laue-Langevin in Grenoble. About 5 g of CoF$_2$ powder sample was put inside a cylindrical Al sample holder that was fixed to the cold tip of the sample stick of a standard orange cryostat. In order to cover an extended $Q$-range, relevant to the present study, and to gain in energy resolution two incident wavelengths of $\lambda_{i}$~= 1.1 and 1.5 {\AA}, respectively, were used. This allowed the Stokes spectrum to be measured at low temperature over a broader dynamical range. The data analysis was done using ILL software tools

The present investigations have been performed with unpolarized neutrons on powder samples of CoF$_2$. The present method is therefore somewhat limited
because it gives only the powder average results and does not give information
of detailed directional Q dependence. Also the unpolarised neutron scattering
cannot distinguish directly whether the scattering is of magnetic or
structural origin. However and since CoF$_2$ contains magnetic Co ions, we can
therefore expect inelastic magnetic scattering due to the interaction of the
neutron spin with the spin and orbital magnetic magnetization of the CoF$_2$
both below and above the magnetic ordering temperature $T_N \approx 39$
K. Also the neutron is sensitive to the structural excitations or phonons
which exist at all temperatures. Thus this technique will measure both
excitations and even highlights their hybrid feature described
before. Magnetic and structural excitations have fortunately different
temperature and Q dependence. Strong and sharp inelastic magnetic excitations
are expected only below the magnetic ordering temperatures. At higher
temperature, the magnetic correlations give broad quasielastic scattering
instead. The intensity of the magnetic excitation peaks decrease with the
momentum transfer Q. The scattering due to structural excitations or phonons
have different temperature and Q dependences. The phonon intensity in general
increases with increase of the momentum transfer Q. Of course intensities of
magnons and phonons are also governed by the magnon and phonon structure
factors and also polarisations which depends on the details of the magnetic
and crystalline structures. However the differences in temperature and Q
dependence for magnons and phonons can be conveniently used for the sake of
identification in order to separate them.

The cross section for neutron scattering by magnetic systems is discussed in details in several text books on neutron scattering \cite{squires78,marshall71}. The expression for neutron scattering cross sections are very complex when the ions have both spin and orbital angular momentum. It can be greatly simplified if the momentum transfer is less than the reciprocal of the radius of the magnetic shell and is given by 
\begin{eqnarray}
\label{magnon}
{\frac{d^2\sigma}{d\Omega d\omega}}  =\frac{k_f}{k_i}\frac{1}{2\pi}\left(\frac{1.91e^2}{2mc^2}\right)^2\\ \nonumber
\times \sum_{ij}|f({\bf Q})|^2\int_{-\infty}^{\infty}\left<{\bf K}(i,0).{\bf K}(j,t)\right>\\ \nonumber
\times \exp \left[i{\bf Q}.({\bf R}(i)-{\bf R}(j))\right]\exp(-i\omega t) dt,
\end{eqnarray}
where $f({\bf Q})$ is the form factor of the ions, ${\bf K}(j,t)$ is the part of the magnetic moment operator, ${\bf L+2S}$, which is perpendicular to ${\bf Q}$, for the ion $i$ ar ${\bf R}(i)$ at the time $t$, ${\bf k}_i$ and ${\bf k}_f$ are the wavevectors of the incident and scattered neutrons. 

The neutron scattering cross section of a phonon mode $({\bf q}j)$ is given by
\begin{eqnarray}
\label{phonon}
{\frac{d^2\sigma}{d\Omega d\omega}}  =\frac{k_f}{k_i}\frac{1}{\omega({\bf q}j)} \\ \nonumber
| \sum_{k}\left(\frac{\hbar}{2M_k}\right)^{\frac{1}{2}}b_k{\bf Q}.{\bf e}(k,{\bf q}j)\\ \nonumber
\times \exp [i({\bf Q-q}).{\bf R}(k)]\exp(-W_k({\bf Q})) | ^2 \\ \nonumber
\times ( \frac{n({\bf q}j)}{n({\bf q}j)+1} )
\end{eqnarray}
where $\exp(-W_k({\bf Q}))$ is the Debye-Waller factor for the $k$th atom, $n({\bf q}j)$ is the population of the mode and is taken for neutron energy loss, $b_k$ is the scattering length of the $k$th atom in the unit cell, $M_k$ its mass, and $e(k,{\bf q}j)$ is the eigen vector in the normal mode $({\bf q}j)$. Note that everything we stated in the previous section about the different Q dependence of the magnons and phonons, follows directly from equations (\ref{magnon}) and (\ref{phonon})

Both magnetic and structural excitations have been investigated \cite{martel68,cowley73} by inelastic neutron scattering on CoF$_2$ single crystals. The magnetic properties of CoF$_2$ are much more complex than those of MnF$_2$ having the same rutile crystal structure and also the same antiferromagnetic structure. Crystal field calculations of Gladney \cite{gladney66} shows that the $L = 3, S = 3/2$ atomic ground state is split by the octahedral crystal field to give an orbital $\Gamma_4$ triplet as the ground state whose properties can be described by an effective $l=1$ operator. The rhombic and tetragonal crystal field distortions and the spin-orbit coupling causes this state to split into six Kramers doublets. The molecular field in the antiferromagnetic phase causes further splitting. Labeling the ground state as A and first three excited states as B,C and D three crystal-field excitations have been labeled \cite{martel68} as A-B, A-C and A-D transitions. Alternatively these states are also labeled \cite{cowley73} as 0, 1, 2, 3. The corresponding excitations are labeled as 0 - 1, 0 - 2 and 0 - 3. The dispersions of these excitations have been measured from single crystal samples along $[100]$ and $[001]$ directions. However, the anisotropy in dispersion of these excitations is found to be rather small. The strongest excitation is A-B which is at about 7.4 meV  for $q = 0$. The other two excitations are A-C and A-D that both lie at about 24-25 meV. 

Fig. \ref{spectra1} shows the temperature evolution of the scattering function $S(Q,\omega)$ measured with incident neutron wavelength of $\lambda = 1.5$ {\AA} and Fig. \ref{spectra2} shows temperature evolution of the scattering function $S(Q,\omega)$ measured with $\lambda = 1.1$ {\AA}. 
At $T = 2$ K we observe two inelastic peaks at about $E = 8$ meV and $E = 24$
meV. The energy of these peaks decreases with increasing temperature. However
whereas the energy of the first peak at $E = 8$ meV becomes almost zero at
$T_N = 38$ K,  the energy of the second peak decreases and then attains a
finite value at $T_N$  and stays constant at about 19 meV at higher
temperatures. Fig. \ref{coffit1} shows the results of the fit of the low-Q
data from CoF$_2$ in the low-energy range in the left two panels. Two Gaussian
peaks have been fitted to account for the elastic peak at zero energy and the
inelastic peak which is at about $E = 8$ meV at $T = 2$ K. The energy of the
inelastic peak decreases with increasing temperature and becomes zero at $T_N
= 38$ K giving rise to quasielastic scattering, which has been fitted with a
Lorentzian function. The results of the fit with Gaussian functions to the
higher-energy peaks are shown in Fig. \ref{coffit2}. At low temperatures below $T_N = 38$ K,  three peaks could be identified and they were fitted with three Gaussian functions. At higher temperatures only a broad peak could be identified and has feen fitted with a single Gaussian function. Fig. \ref{Tdep} (a) shows the temperature variation of the energy and the half-width at half-maximum (HWHM) of the low-energy inelastic magnetic peak of CoF$_2$. In Fig. \ref{Tdep} (b) the temperature variation of the energy of the higher-energy inelastic magnetic peak of CoF$_2$ has been shown. The temperature variation of the low energy inelastic peak which is at about 8 meV at 2 K  shows clearly that this peak is due to magnetic excitations. The energy of this peak decreases continuously and becomes almost zero at $T_N = 38$ K giving rise to quasielastic scattering of Lorentzian form. The higher energy peak at about 24 meV on the other hand seems to be a hybrid peak due to the magnon-phonon coupling. The peak persists at temperature higher then $T_N$ and its energy remains constant at about 19 meV. 

Fig. \ref{Qdep} shows the Q dependence of the intensity for the  inelastic
peaks at $E = 7.3, 14.0$, $23.5$, $31.6$ and $43.0$ meV at T = 2, 20, 40, 70
and  100 K. Fig.  \ref{Qdep} (a) shows the Q dependence of the peak at E = 7.3
meV for several temperatures. The intensity of the peak at $E = 7.3$ meV at T
= 2 K decreases continuously with increasing Q and then becomes more or less
constant at higher Q. This indicates that the inelastic peak is of mostly
magnetic origin. This is also supported by its intensity and Q dependence at
$T = 100$ K. However even at T = 100 K the intensity is not zero. This
suggests that there exists some phonon contribution to this peak as well. The
Q dependence at T = 100 K shows that the intensity increases slightly at
higher Q and this also confirms that there exist some phonon contribution to
this peak.  Fig.  \ref{Qdep} (b) shows the Q dependence of the inelastic peak
at E = 14 meV at several temperatures.  The intensity of this peak first
decreases with increasing Q and then increases very much at higher Q. This
suggests that the peak has a hybrid phonon and magnon character. The Q
dependence of such a hybrid peak is also expected to be more
exotic. Fig. \ref{Qdep} (c) shows the Q dependence of the inelastic peak at E
= 23.5 meV. Its Q dependence is very similar to the peak at E = 7.3 meV shown
in Fig.  \ref{Qdep} (a). Even its temperature dependence is very similar. The
intensity of the peak at all Q is higher at lower temperature. The decrease of
its intensity with increasing Q suggests that it is magnon like. However like
the peak at E = 7.3 meV the peak at E = 23.5 also shows slight tendency of
increasing in intensity at higher Q suggesting that the peak has some phonon
like character.  Fig. \ref{Qdep} (d) shows the Q dependence of the inelastic
peak at E = 31.6 meV. It looks very similar to that at E = 14 meV and consists
clearly of both phonon and magnon contributions. Lastly the Q dependence of
the inelastic peak shown in Fig. \ref{Qdep} (e) shows that this peak is almost
pure phonon like because its intensity increases with Q and temperature. We have fitted the Q dependence of the intensity for the peaks at 14 and 31.6 meV assuming them to consist of both magnon and phonon contributions and also assuming that the Q dependence has oscillatory cosine wave form due to the periodicity of the lattice. We fitted the intensity by the equation
\begin{equation}
I(Q)=a_1+a_2Q+a_3Q^2+a_4\cos\left(\frac{2\pi}{a_5}Q+a_6\right)
\end{equation}
which consists of a polynomial plus an oscillatory cosine function. Here $a_1, a_2$, and $a_3$ are coefficients of a second order polynomial and $a_4$ is the amplitude and $a_5$ is the periodicity of the oscillatory part superimposed on the polynomial function and $a_6$ is the phase.
This is a much simplified function but since it contains term proportional to $Q$ describing Q dependence of the magnetic intensity  and a term proportional to $Q^2$ describing the $Q$ dependence of the phonon intensity and has also a superimposed oscillatory term describing the effect of lattice periodicity, it describes the intensity variation rather well. The fitted periodicity is in agreement with the average lattice periodicity. Since the fits for different temperature in Fig. \ref{Qdep} (b) is not clearly visible we show a typical fit at T = 2 K separately in Fig.   \ref{Qdep} (f). The fit looks rather convincing indeed. In fact the Q dependences of the intensities for all five energies shown in Fig. \ref{Qdep} (a-f) show oscillations but the statistics and also the resolution of the data shown in Fig. \ref{Qdep} (a), \ref{Qdep} (c) and \ref{Qdep} (e) are not good enough for reasonable fits. 

In conclusion we investigated the phonon and spin dynamics of CoF$_2$ by inelastic neutron scattering on powder samples by time-of-flight neutron spectroscopy with unpolarized neutrons. The temperature and the Q dependence of the intensity of the inelastic signals enabled us to identify their phonon, magnon or hybrid origin at least to a first approximation. We note that there exist considerable phonon-magnon interaction in this simple antiferromagnetic system also evidenced by the strong magnetoelastic coupling observed by neutron diffraction \cite{chatterji10a}. However, in order to be certain about the hybrid nature of these inelastic peaks and evaluate quantitatively phonon-magnon interaction strength, polarized neutron scattering investigation is desirable.

\end{document}